\documentclass{ptephy_v1}
\preprintnumber{XXXX-XXXX} 

\begin{document}

\title{First reconstruction of absolute three-dimensional position of nuclear recoils using a negative ion $\mu$-TPC for dark matter search experiments}

\author[1]{Satoshi Higashino}
\author[1]{Takuya Shimada}
\author[2]{Tomonori Ikeda}
\author[1]{Hirohisa Ishiura}
\author[1]{Ryo Kubota}
\author[1]{Ayaka Nakayama}
\author[1]{Mizuno Ofuji}
\author[1]{Kentaro Miuchi}

\affil{Department of Physics, Graduate School of Science, Kobe University, Rokkodai-cho, Nada-ku, Kobe-shi, Hyogo, 657-8501, Japan \email{higashino@people.kobe-u.ac.jp}}
\affil{Division of Physics and Astronomy, Graduate School of Science, Kyoto University, Kitashirakawaoiwake-cho, Sakyo-ku, Kyoto-shi, Kyoto, 606-8502, Japan}

\begin{abstract}
Studies of micro time projection chambers using negative ion gas (NI$\mu$TPC) are conducted especially for direction-sensitive dark matter searches.
A NI$\mu$TPC enables to measure the absolute position in the drift direction for self-triggering TPCs.
This study provides a development of a NI$\mu$TPC using a micro pixel chamber ($\mu$-PIC) with dedicated readout electronics, and an evaluation of the capability of three-dimensional absolute position reconstruction of nuclear recoil using a neutron source.
The absolute track position was reconstructed throughout the drift volume with an efficiency of 70~$\pm$~5\%.
This work marks an important step for the NI$\mu$TPCs towards the practical use for directional dark matter searches.
\end{abstract}

\subjectindex{Dark matter, $\mu$TPC, Negative ion, SF$_{6}$}

\maketitle

\section{Introduction}
\label{sec:intro}

For the decades, dark matter searches have been taken place by many experiments although nature is yet to be cleared.
A direct dark matter search, which detect nuclear recoil by the dark matter, is one of the useful methods to explore its nature.
Especially, assuming that a dark matter is Weakly Interacting Massive Particle (WIMP), direct dark matter searches make the strongest limits~\cite{Arbey:2021gdg}.
One of the methods of the dark matter searches is to measure the deposited energy of recoil nuclei and evaluate any excesses over the background-only hypothesis in the energy spectra.
On the other hand, the annual modulation and the directionality would be ones of more characteristic signatures of WIMPs.
The annual modulation of WIMP-induced event rate comes from the orbital motion of the Earth around the sun, while the directionality of the signals is made by the rotation of the solar system around the Milky Way Galaxy.
While the magnitude of the annual modulation could be a few percent~\cite{Baum:2018ekm}, the forward-backward ratio of nuclear recoil events in the rotation direction could be an order of magnitude~\cite{Spergel:1987kx}.
Moreover, the directional method would distinguish WIMPs from neutrino-induced background (CE$\nu$NS), which allows to explore beyond the neutrino floor~\cite{Billard:2013qya}.
Therefore the directionality measurement is a significant tool to reveal the nature of WIMP dark matter.

One of the widely-used methods of direction-sensitive direct dark matter search is to detect trajectories of recoil nuclei using a gaseous time projection chamber (TPC).
Several experiments such as DRIFT~\cite{DRIFT:2016utn} or NEWAGE~\cite{Ikeda:2021ckk} have carried out dark matter searches in underground observatories using their own gaseous TPCs.
While NEWAGE uses CF$_{4}$ gas for the TPC volume and target for the dark matters, DRIFT uses CS$_{2}$ gas in the chamber.
The CS$_{2}$ gas is so-called a "negative ion gas", which has electro-negative characteristics.
In the negative ion gases, ionized electrons are captured by their own molecules right after their production and negative ions, instead of electrons, drift in the gas volume suffering less diffusions.
Thus it could improve the precision of track reconstruction.
Furthermore, it turned out that negative ion gas such as CS$_{2}$ creates "minority carriers" in the electron capture process.
The creation of minority carrier allows us to reconstruct tracks with their absolute coordinate in drift direction~\cite{DRIFT:2014bny}.

After the "discovery" of the CS$_{2}$ gas, SF$_{6}$, which is safer than CS$_{2}$ and is suitable for the spin-dependent search using fluorine (F) recoils, was found to behave similarly~\cite{Phan:2016veo}.
Although several studies on SF$_{6}$ gas had been conducted~\cite{Ikeda:2017jvy,Baracchini:2017ysg,Ligtenberg:2021viw} aiming for dark matter searches, no practical TPCs with SF$_{6}$ gas had been developed because of the difficulty of the development of micro-patterned gaseous detectors (MPGDs) with suitable readout electronics for negative ion TPCs. 
Recently, a prototype negative ion micro TPC (NI$\mu$TPC) was developed with a MPGD with dedicated electronics~\cite{Ikeda:2020pex}, which demonstrated the first three-dimensional tracking of alpha particles with absolute coordinate reconstructions.
This paper describes further studies on the NI$\mu$TPC performance of three-dimensional track reconstructions of nuclear recoils using a neutron source.
We developed a new preamplifier ASIC with low noise and implemented a self-triggering system in the back-end electronics.
The development of the NI$\mu$TPC made a portal of dark matter searches using negative ion gases.

The rest of this paper consists of three parts. Section~\ref{sec:detector} describes the NI$\mu$TPC together with its data acquisition system.
Section~\ref{sec:experiment} shows experimental results using alpha and neutron sources.
We conclude our studies in Section~\ref{sec:conclusion}.

\section{Detector}
\label{sec:detector}

In this section we describe the detector including the data acquisition system for evaluations as a device for a dark matter search.
This section also explains the behavior of SF$_{6}$ gas in the TPC.

\subsection{Negative ion micro TPC}
\label{sec:NIuTPC}

\begin{figure}[h]
    \centering
    \includegraphics[width=0.8\textwidth]{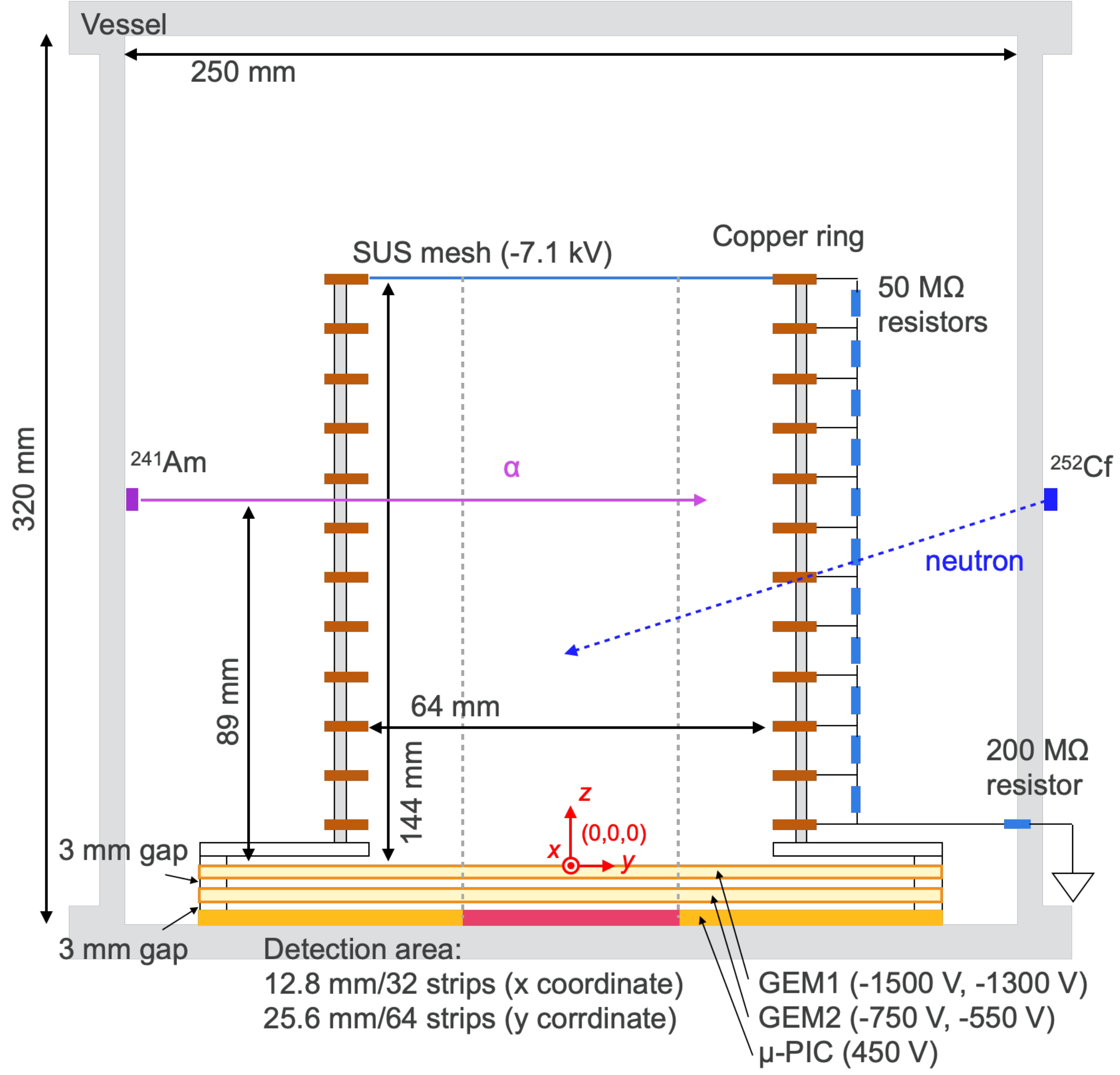}
    \caption{Schematic of the NI$\mu$TPC.}
    \label{fig:NIuTPC}
\end{figure}

Figure~\ref{fig:NIuTPC} shows the schematics of our NI$\mu$TPC.
We used one of the MPGD variations, a micro pixel chamber ($\mu$-PIC~\cite{Hashimoto:2020xly}; Dai Nippon Printing Co., Ltd.), as a readout device.
It has 256~$\times$~256~pixels with 400~$\mu$m pitch in a region of 10~$\times$~10~cm$^{2}$.
These electrodes are connected by 256~anode strips and 256~cathode strips, which cross orthogonal each other. 
It enables to detect two-dimensional positions where a charged particle deposits its energy.
In this study, we used only 32~anode strips and 64~cathode strips placed in the center of the $\mu$-PIC due to the limited readout channels of electronics.
The anode strips have cylindrical electrodes with a diameter of 60~$\mu$m, and the cathode strips have circular openings with a diameter of 250~$\mu$m surrounding each anode electrode.
Such a pixel structure forms strong electrical field around anode electrodes and it makes an avalanche and consequently enables to amplify the electrons in the gas.
An additional gas amplification is made by two 10~$\times$~10 cm$^{2}$ layers of gas electron multipliers (GEM~\cite{Sauli:2016eeu}; SciEnergy Co., Ltd.) above the $\mu$-PIC, with 3~mm transfer gaps.
The GEM is a 100~$\mu$m thick liquid crystal polymer with 5~$\mu$m copper coatings on both sides.
The GEM has 70~$\mu$m holes with a pitch of 140~$\mu$m in entire surface~\cite{Tamagawa:2009xa}.
The TPC volume with a drift length of 144~mm is formed with 12 copper rings with an inner diameter of 64~mm.
The rings are connected via 50~M$\Omega$ registers to form a uniform electrical field within the TPC volume.
The drift plane was made of a stainless-steel mesh.
Voltages of -7.1~kV, -1500~V, -1300~V, -750~V, -550~V, 0~V, and 450~V are supplied to the drift plane, top and bottom copper layers of two GEMs (four layers in total), and cathode and anode electrodes of the $\mu$-PIC, respectively.
The total gas gain was 2800 in the condition above and the electrical field of the detection volume was 0.40~kV/cm.

The outer vessel was made of stainless-still and was filled with pure SF$_{6}$ gas at 20~Torr.
The gas was circulated at a rate of 0.1~L/min.
A contamination of water vapor is filtered-out by Zeolum (A-3) installed in the circulation system at the room temperature.
The amount of the water contamination was monitored with a dew point meter (Easidew Transmitter, MICHELL Instruments Ltd.).
During the measurement, the water contamination level was kept below 2500~ppm.

\subsection{Data acquisition system}
\label{sec:DAQ}

\begin{figure}[h]
    \centering
    \includegraphics[width=1.0\textwidth]{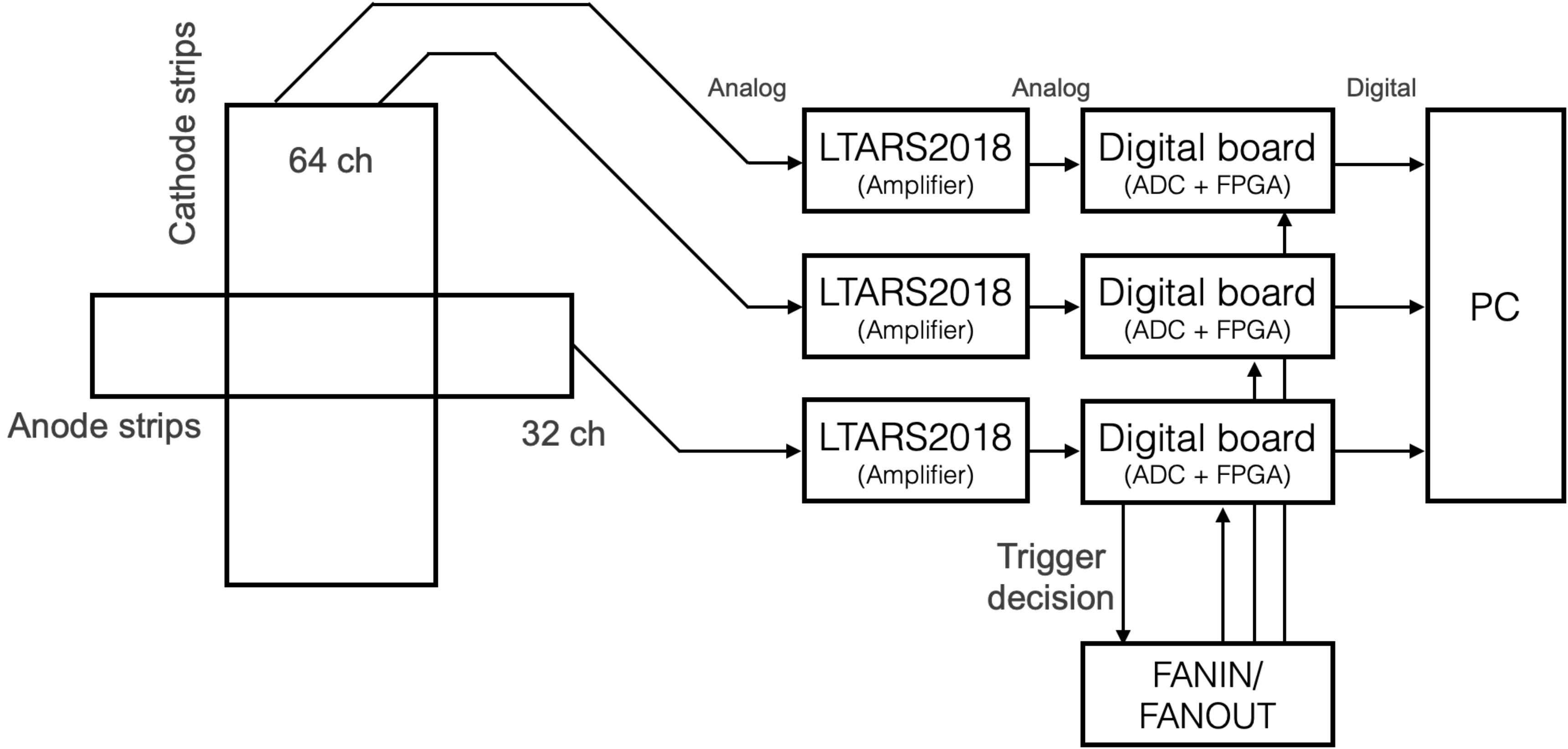}
    \caption{Diagram of the data acquisition system.}
    \label{fig:DAQ}
\end{figure}

Figure~\ref{fig:DAQ} shows the schematics of the data acquisition (DAQ) system.
Each electrode was AC-coupled with a 100~pF capacitor to its amplifier.
The charge was amplified by a dedicated preamplifier ASIC (LTARS2018~\cite{Kishishita:2020skm}), updates of LTARS2014~\cite{Sakashita:2015qqa} and LTARS2016~\cite{Nakazawa:2019hvo}.
Since the previous versions had a problem of the electrical noise, the latest version LTARS2018 improved this issue.
The LTARS2018 allows to readout 16 strips per chip, and we developed a printed circuit board loaded with two LTARS2018 chips.
Due to limited numbers of available boards, one board for anode strips (32~channels) and two for cathode (64~channels) are used in this study.
Amplified signals were digitized at general purpose digitization boards.
The digitization board mainly consists of 2.5~MHz sampling 12~bit ADC with a dynamic range of 2~V, Xilinx Artix-7 FPGA, NIM I/O ports and Ethernet interface.
The digitized data were sent to a computer using the SiTCP technology~\cite{Uchida:2007cwb}.
The self-triggering algorithm was developed and implemented to the firmware for this study.
At least three channels detecting signals with a certain threshold were required to issue triggers in this study.

\subsection{Characteristics of SF$_{6}$ gas in TPC}
\label{sec:SF6}

\begin{figure}[htbp]
	\begin{center}
		\begin{tabular}{c}
			\begin{minipage}{0.57 \hsize}
				\begin{center}
					\includegraphics[clip, width=8.0cm]{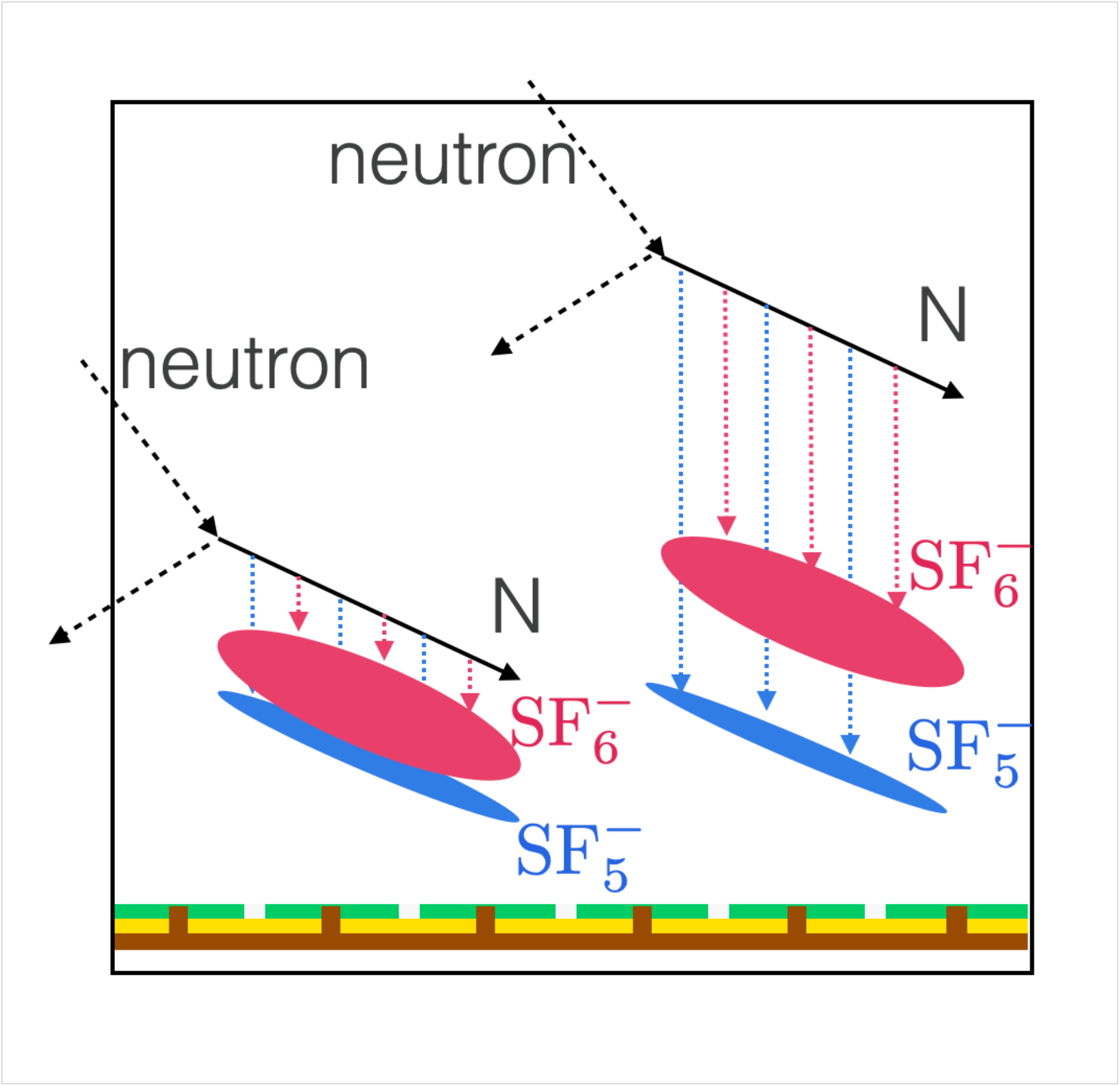}
					\hspace{1.6cm}
				\end{center}
			\end{minipage}
			\begin{minipage}{0.43 \hsize}
				\begin{center}
					\includegraphics[clip, width=6.4cm]{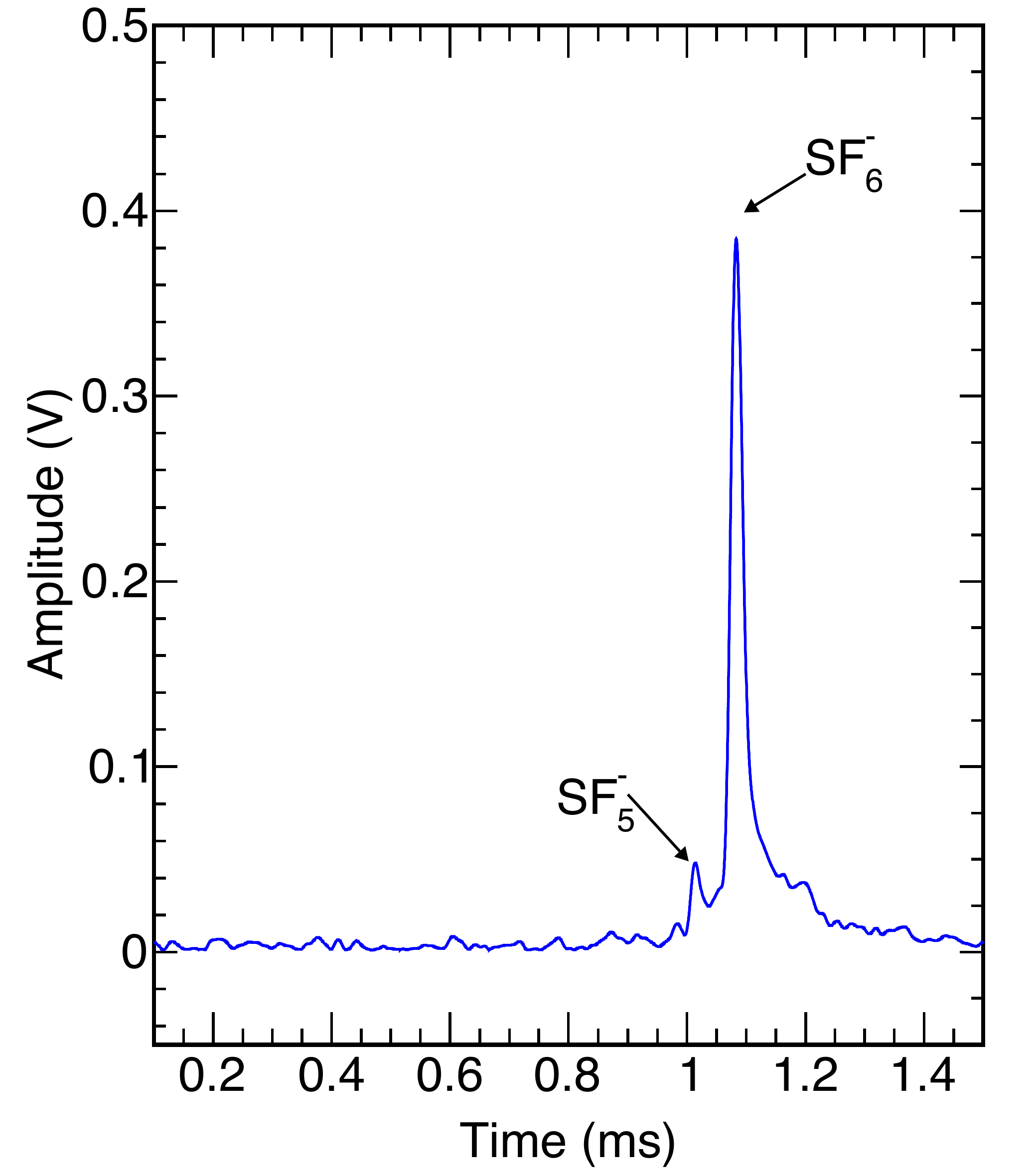}
					\hspace{1.6cm}
				\end{center}
			\end{minipage}
		\end{tabular}
		\caption{Illustration of SF$_{6}^{-}$ and SF$_{5}^{-}$ ion drifts (left), and a typical pulse of a TPC with SF$_{6}$ gas (right)~\cite{Phan:2016veo}.}
		\label{fig:sf6_property}
	\end{center}
\end{figure}

The property of the SF$_{6}$ gas is described in reference~\cite{Phan:2016veo}.
After the ionization caused by a charged particle, ionized electrons are immediately attached to SF$_{6}$ molecules and then create metastable excited products of SF$_{6}^{-*}$ ions.
This metastable SF$_{6}^{-*}$ ion then forms a negative ion of SF$_{6}^{-}$ or SF$_{5}^{-}$.
The branching fraction of SF$_{6}^{-}$ and SF$_{5}^{-}$ is known to be 97\% and 3\%, respectively~\cite{Phan:2016veo} in a room temperature.
Figure~\ref{fig:sf6_property} illustrates the behavior of the drift of SF$_{6}^{-}$ and SF$_{5}^{-}$.
Due to the difference of their masses, the drift velocity of SF$_{6}^{-}$ is less than that of SF$_{5}^{-}$, and consequently two pulses can appear with a time difference depending on the drift length.
The right plot of Fig.~\ref{fig:sf6_property} shows such a characteristic signal.

\section{Experiment and result}
\label{sec:experiment}

In this section, we describe the detector calibration, the measurement, and its results.
Dark matter induced signals were emulated by nuclear recoil events caused by neutrons from a $^{252}$Cf source.
Before the measurement, an energy calibration was taken place.

\subsection{Energy calibration}
\label{sec:calibration}

\begin{figure}[h]
    \centering
    \includegraphics[width=1.0\textwidth]{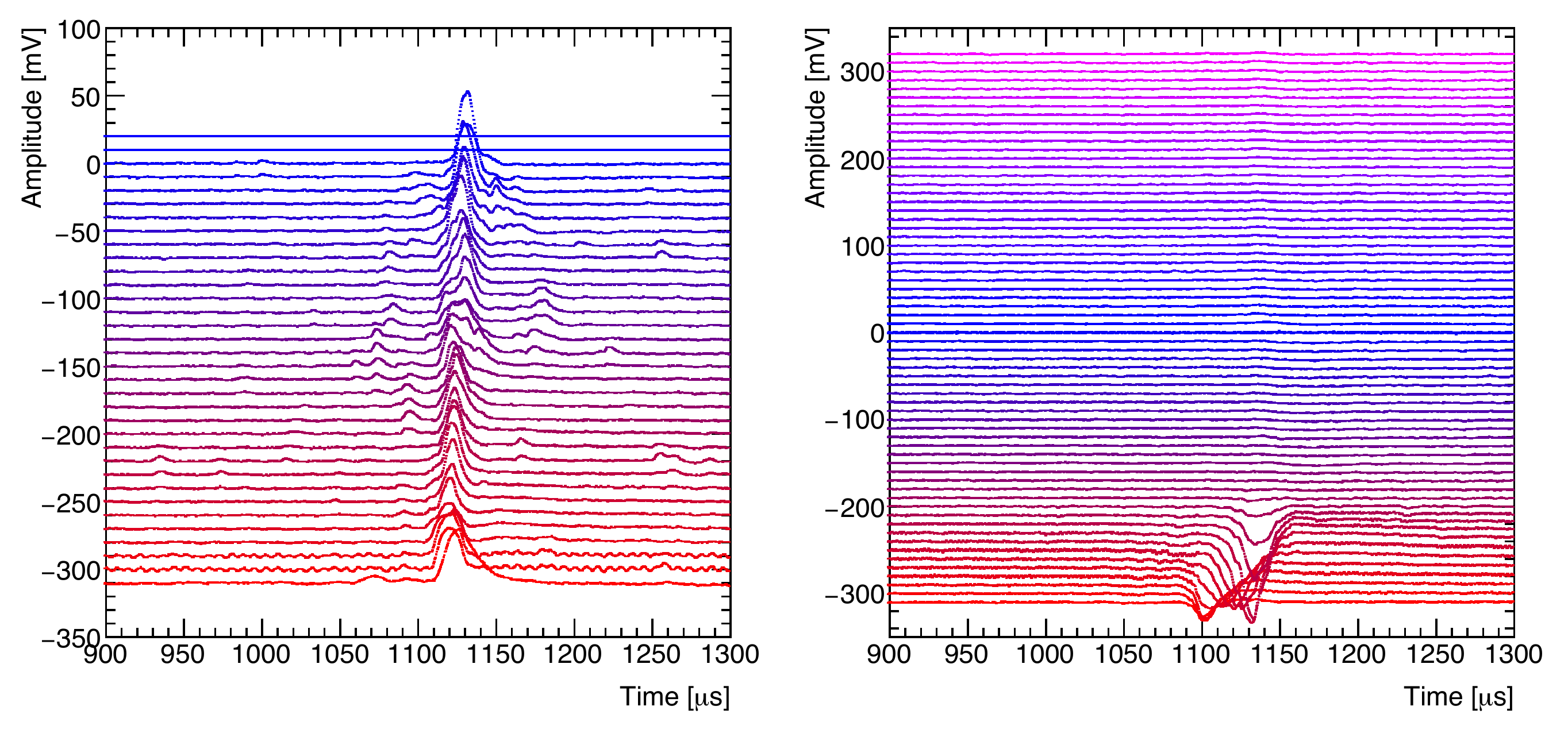}
    \caption{Waveforms of a typical alpha-ray event recorded through the anode (left) and cathode (right) strips.}
    \label{fig:waveform_alpha}
\end{figure}

\begin{figure}[htbp]
	\begin{center}
		\begin{tabular}{c}
			\begin{minipage}{0.5 \hsize}
				\begin{center}
					\includegraphics[clip, width=7.4cm]{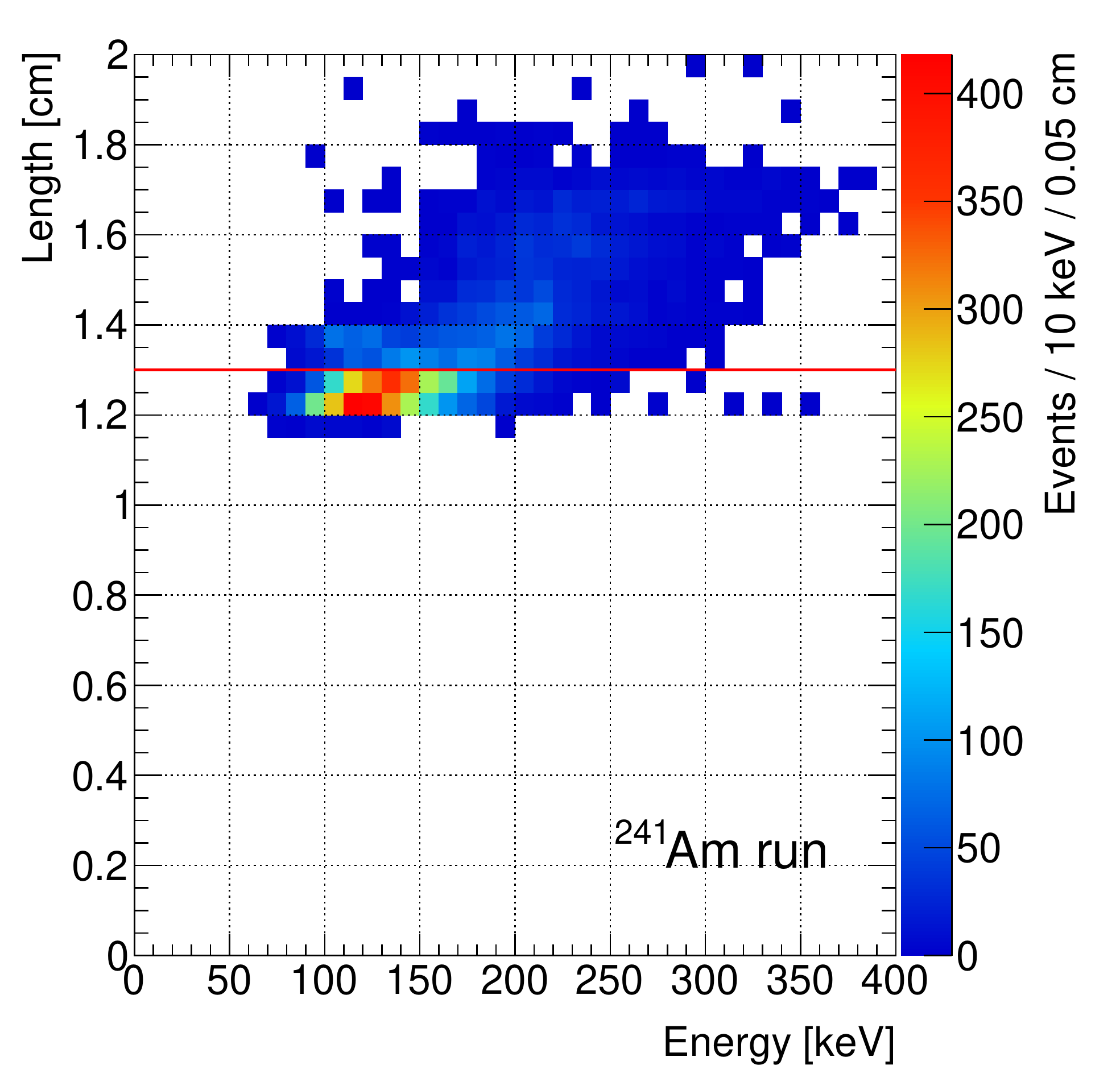}
					\hspace{1.6cm}
				\end{center}
			\end{minipage}
			\begin{minipage}{0.5 \hsize}
				\begin{center}
					\includegraphics[clip, width=7.4cm]{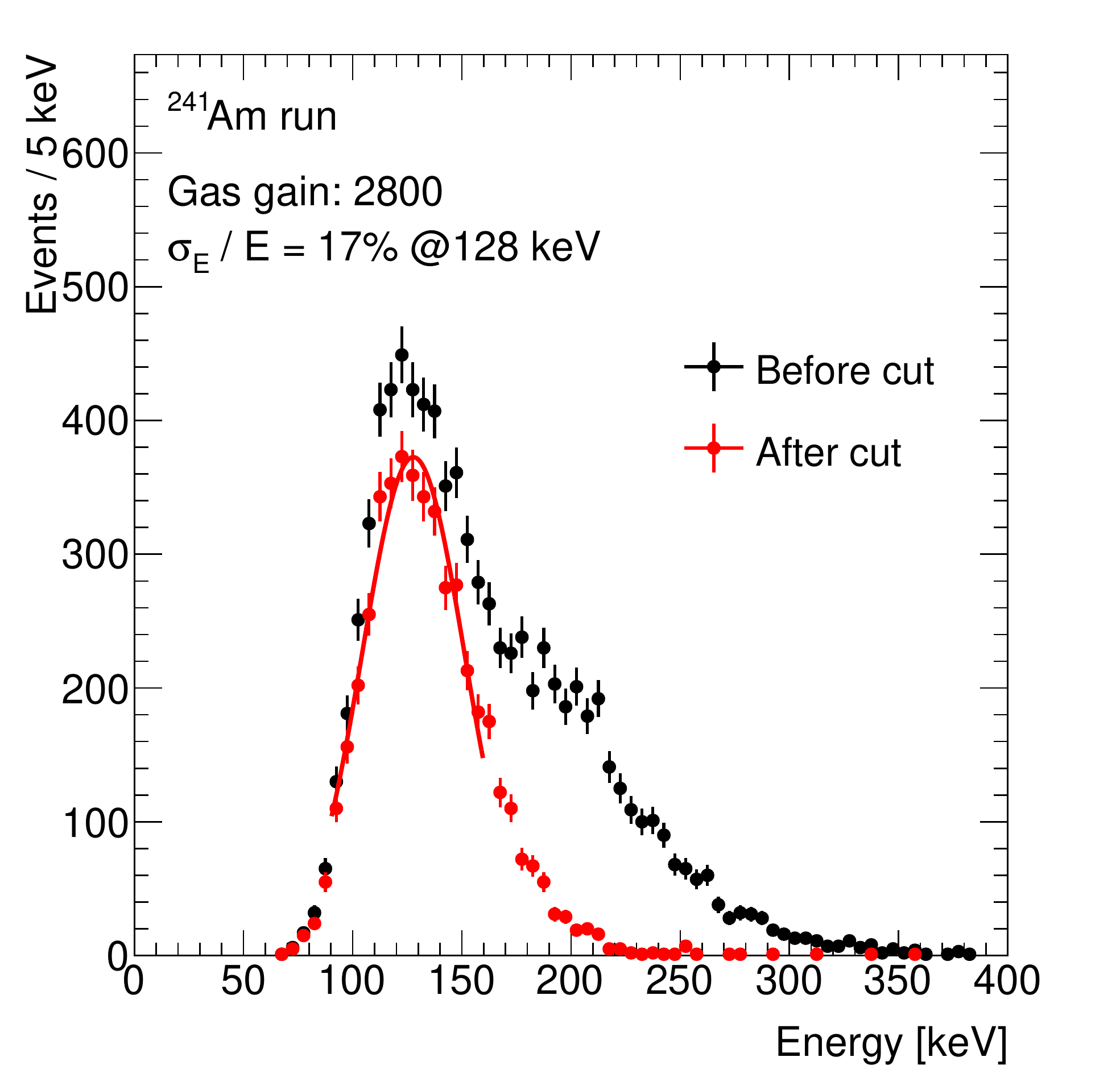}
					\hspace{1.6cm}
				\end{center}
			\end{minipage}
		\end{tabular}
		\caption{Distribution of deposited energy and track length (left) and its projection for deposited energy before and after the length cut described in the text (right). The red line in the left plot shows the selection line of alpha ray events which pass through the detection region perpendicular to the anode strips.}
		\label{fig:cal_alpha}
	\end{center}
\end{figure}

The energy calibration was performed using an $^{241}$Am source which emits 5.5~MeV alpha rays.
The $^{241}$Am source was placed inside the chamber.
Figure~\ref{fig:waveform_alpha} shows a typical waveform of an alpha-ray event for each anode strip.
Here both SF$_{6}^{-}$ and SF$_{5}^{-}$ peaks are clearly seen in anode strips.
The energy deposition were measured by integrating the ADC values of SF${_{6}^{-}}$ pulses, and calibrated so that the peak value of the ADC distribution is corrected to be 128~keV which is obtained by Geant4 simulation~\cite{GEANT4:2002zbu,Allison:2006ve,Allison:2016lfl}.
A track was reconstructed by positions of hit strips in anode and cathode, and their hit timing defined with a certain threshold.
The left plot of Fig.~\ref{fig:cal_alpha} shows the two-dimensional distribution of the deposited energy and the track length.
The tail component shown in the higher energy side, which is due to the events passing through the detection region diagonally, was cut by selecting events with length less than 1.3~cm.
The right plot of Fig.~\ref{fig:cal_alpha} shows the projection for the deposited energy.
The energy resolution was calculated by the distribution after the length cut.
It was 17\% in sigma at the energy of 128~keV. 
In this paper, the following discussions are based on the energy calibrated with alpha rays.

\subsection{Nuclear recoil measurement}
\label{sec:NRrun}

The detection capability of nuclear recoil events was evaluated with a $^{252}$Cf neutron source.
Measurements were carried out with a source placed at (0~mm, 133~mm, 89~mm) position for the DAQ time of 447~minutes.
Energy calibrations were continuously taken place with an interval of about 120~min.
The gas gain, gas pressure, dew point, voltages and currents of GEMs and $\mu$-PIC were monitored and were stable during the experiment.

\begin{figure}[h]
    \centering
    \includegraphics[width=0.8\textwidth]{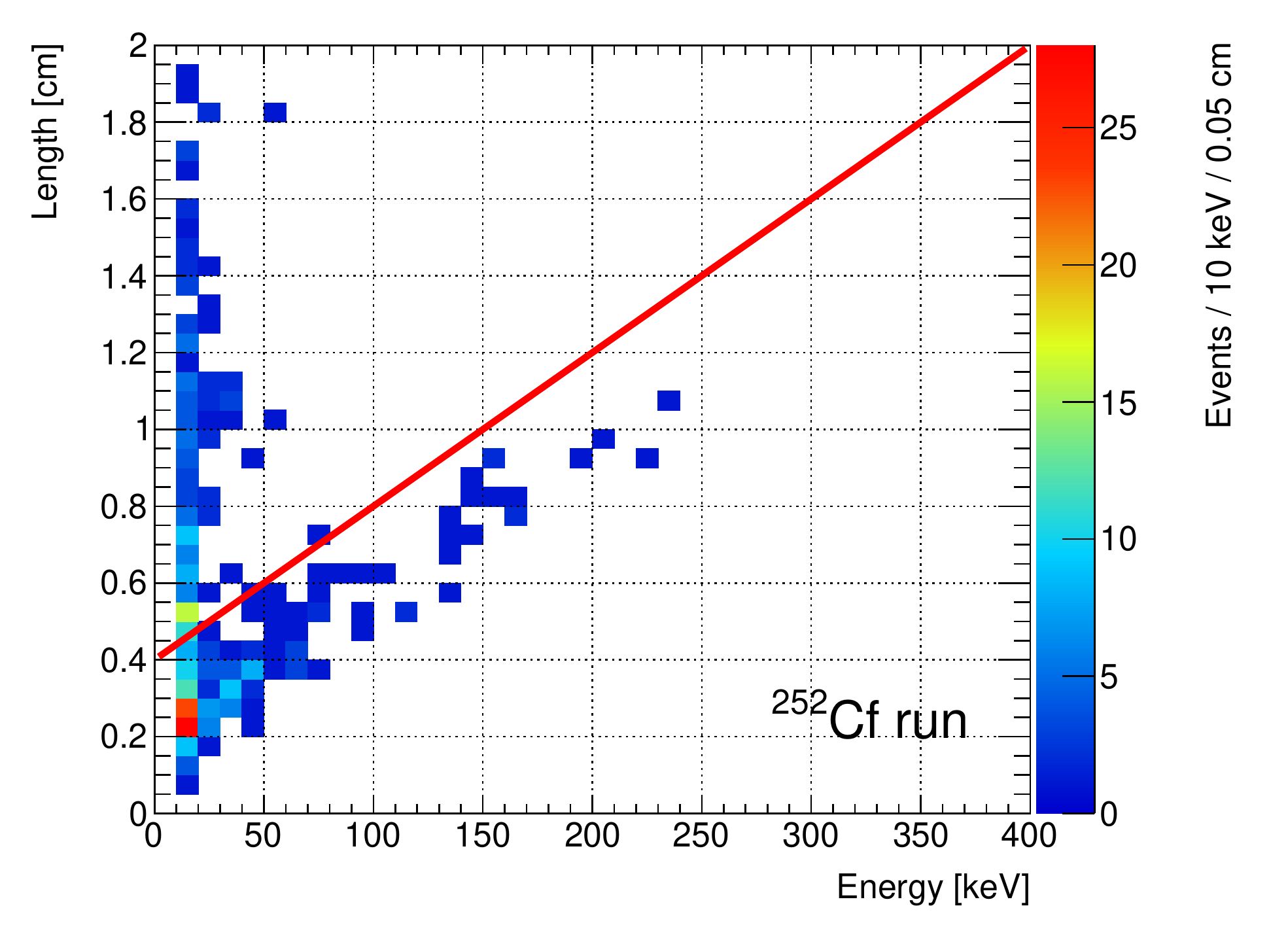}
    \caption{Distribution of the energy and track length in $^{252}$Cf run after the first cut. Red line shows the cut line for the energy-length selection. Events below the red line were selected.}
    \label{fig:enelen_nr}
\end{figure}

In order to reduce background events such as alpha rays or electron recoils caused by ambient gammas, event selections described below were applied.
First, events with signals at the edges of the detection region were rejected to reduce alpha-ray backgrounds.
Figure~\ref{fig:enelen_nr} shows the distribution of energy and track length after the first cut.
The distribution shows two components: one is the events with correlation between energy and length, and the other is non-correlated ones almost along with $y$ axis.
The correlated events are nuclear recoils caused by neutrons, and others are mainly electron recoil backgrounds induced by ambient gamma rays.
To reduce electron recoil backgrounds, we selected events below the red line drawn in Fig.~\ref{fig:enelen_nr} as a second event selection.

\begin{figure}[h]
    \centering
    \includegraphics[width=1.0\textwidth]{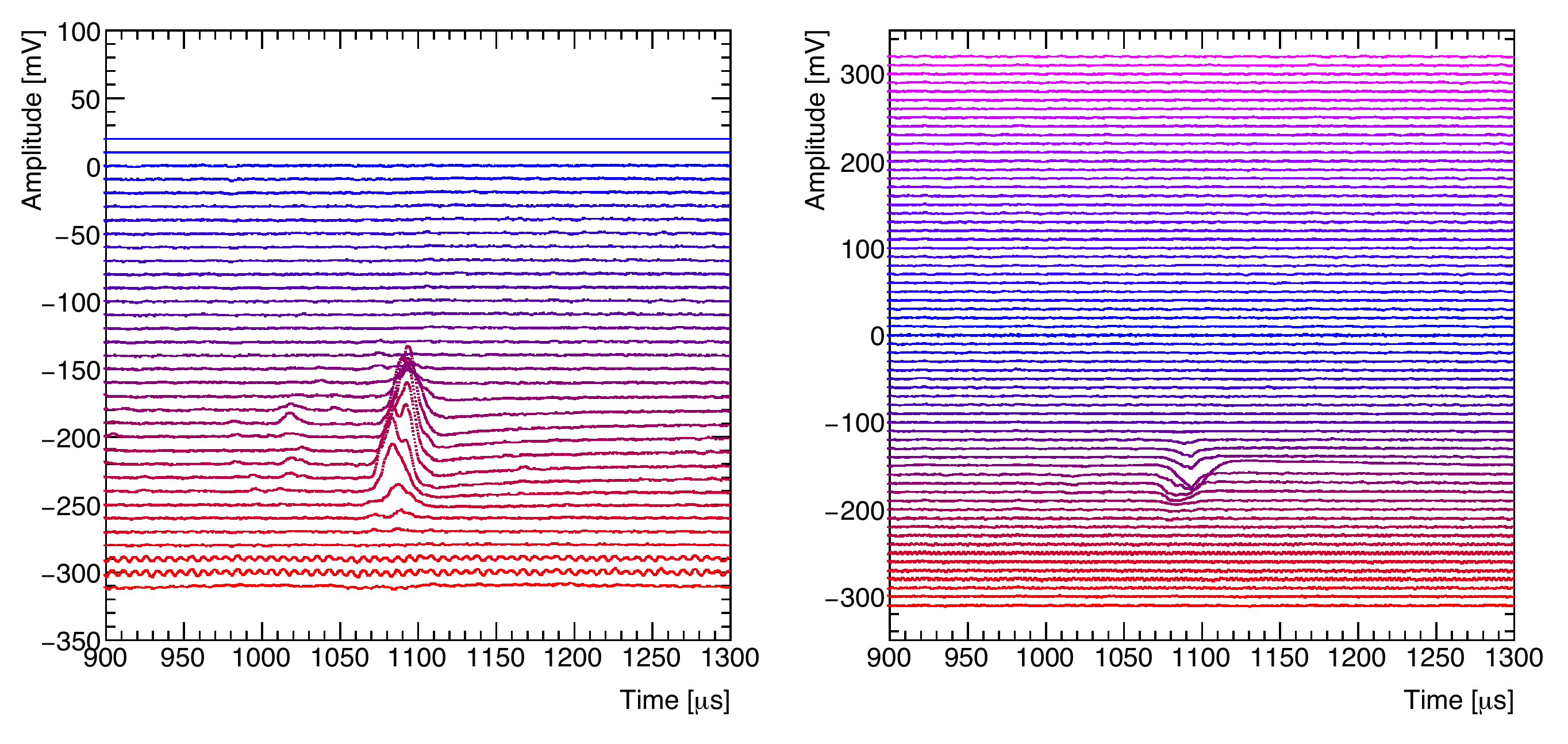}
    \caption{Waveforms of a typical nuclear recoil event in anode (left) and cathode (right) strips.}
    \label{fig:waveform_nr}
\end{figure}

Figure~\ref{fig:waveform_nr} shows a typical waveform of a nuclear recoil event which passed through the selections above.
In this event, both SF$_{5}^{-}$ and SF$_{6}^{-}$ signals are visible in the time of around 1020~$\mu$s and 1090~$\mu$s, respectively, in anode strips.
Signals in cathode strips are relatively smaller than those from anode strips.
This issue is specific to the $\mu$-PIC readout and it is related to the positive ion backflow.

\begin{figure}[h]
    \centering
    \includegraphics[width=0.8\textwidth]{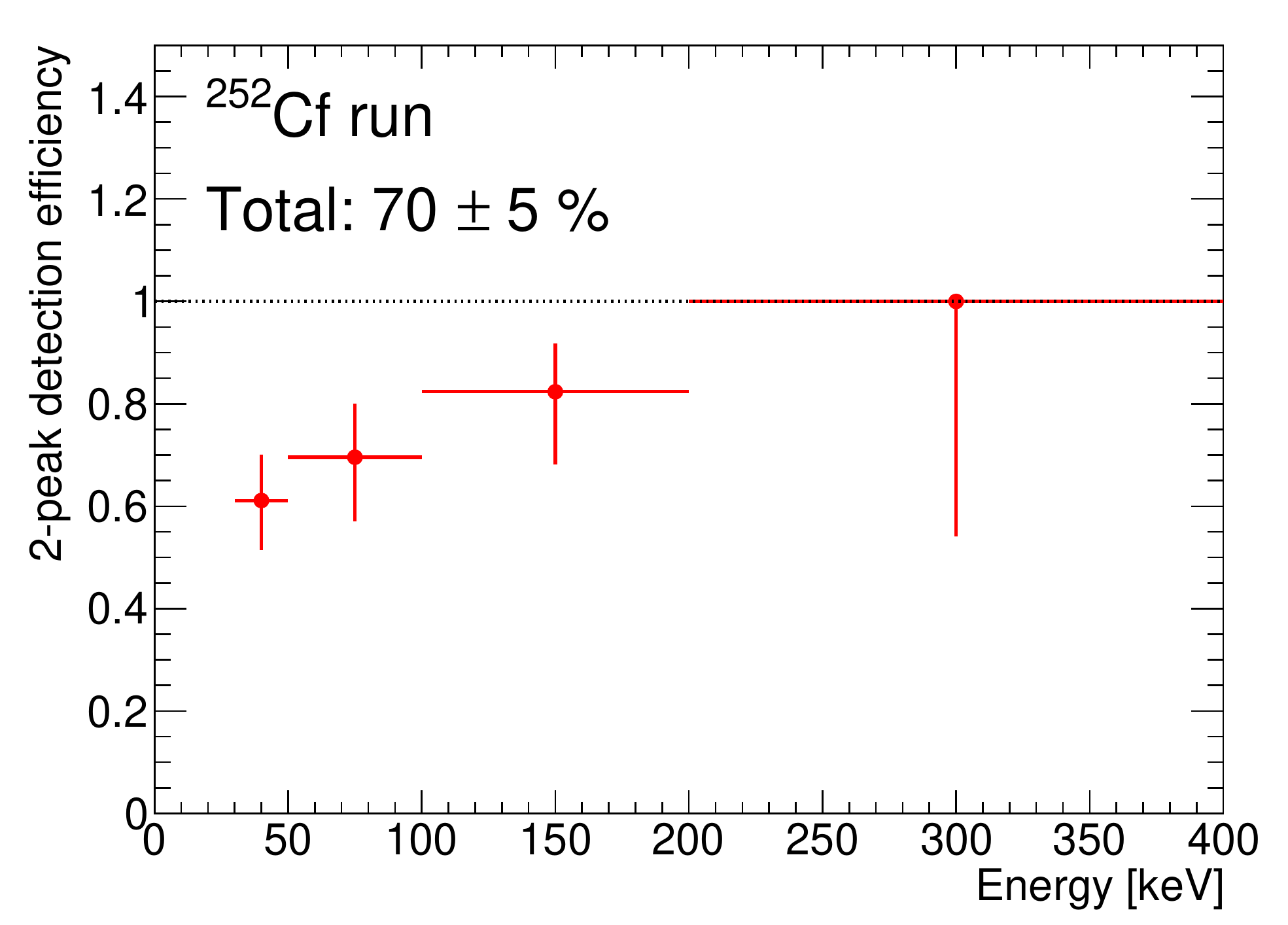}
    \caption{Detection efficiencies of SF$_{5}^{-}$ peaks as a function of energy measured in the $^{252}$Cf run.}
    \label{fig:minorityEff}
\end{figure}

The detection efficiency of SF$_{5}^{-}$ peaks was evaluated for the events passing all selections.
The definition of SF$_{5}^{-}$ peak detection is described here:
\begin{itemize}
\item The search region of interest was defined as [-180~$\mu$s, -30~$\mu$s] from the highest peak position for each channel.  
\item The highest peak above a certain threshold in the region of interest was defined as the SF$_{5}^{-}$ peak.
\item At least one channel was required to define events with a SF$_{5}^{-}$ peak.
\end{itemize}  
Figure~\ref{fig:minorityEff} shows the detection efficiencies of SF$_{5}^{-}$ peaks for each energy slice.
The overall efficiency reached 70~$\pm$~5\%, which was enough to apply this detector for dark matter searches.
There are two reason for the efficiency loss.
One is due to the lack of deposited energy and consequently SF$_{5}^{-}$ peaks are below the threshold, especially in the lower energy range.
This issue will be improved by the increase of the gas gain.
The other is that SF$_{5}^{-}$ peak is merged into the SF$_{6}^{-}$ peak if a nuclear recoil occurs near the GEMs + $\mu$-PIC system.
This is not only for the case of lower energy but also for higher energy events.
Since $\Delta t$ was required to be greater than 30~$\mu$s, the acceptance is estimated to be 80\% by taking the solid angle from the $^{252}$Cf source into account.
Therefore we obtained reasonable efficiencies in the energy range of [100~keV, 400~keV].

\begin{figure}[htbp]
	\begin{center}
		\begin{tabular}{c}
			\begin{minipage}{0.5 \hsize}
				\begin{center}
					\includegraphics[clip, width=7.4cm]{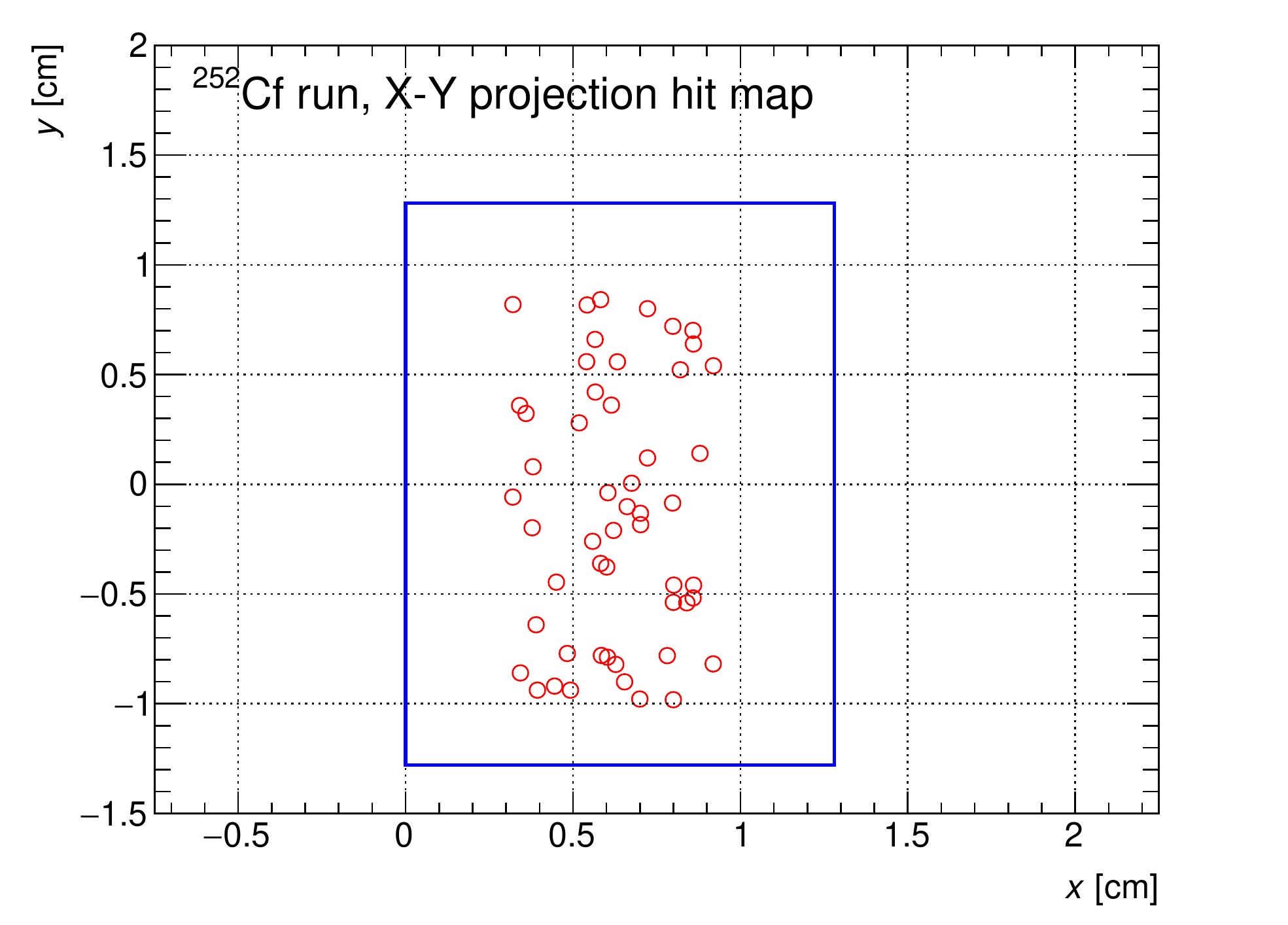}
					\hspace{1.6cm}
				\end{center}
			\end{minipage}
			\begin{minipage}{0.5 \hsize}
				\begin{center}
					\includegraphics[clip, width=7.4cm]{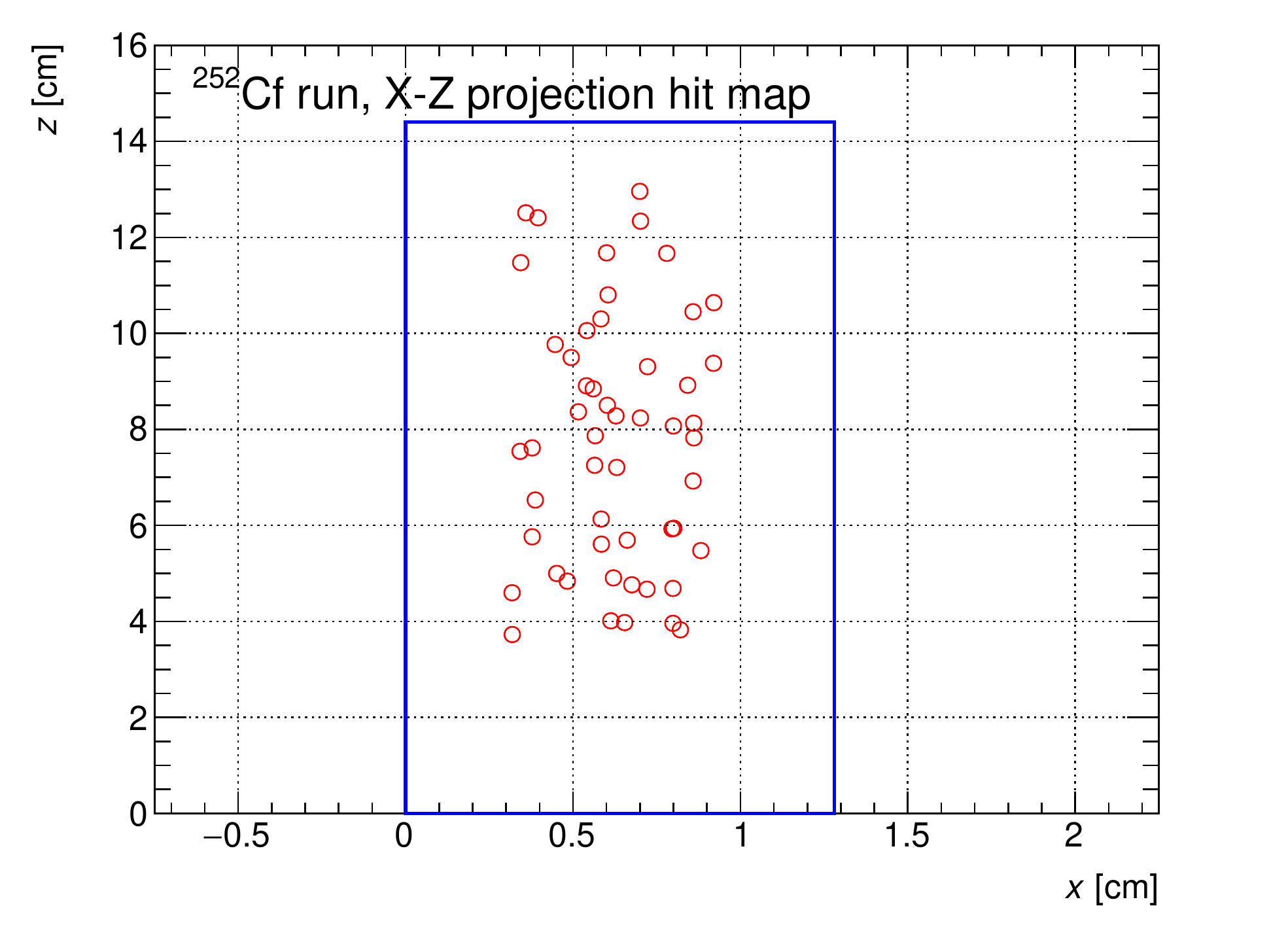}
					\hspace{1.6cm}
				\end{center}
			\end{minipage}
		\end{tabular}
        \begin{tabular}{c}
			\begin{minipage}{0.5 \hsize}
				\begin{center}
					\includegraphics[clip, width=7.4cm]{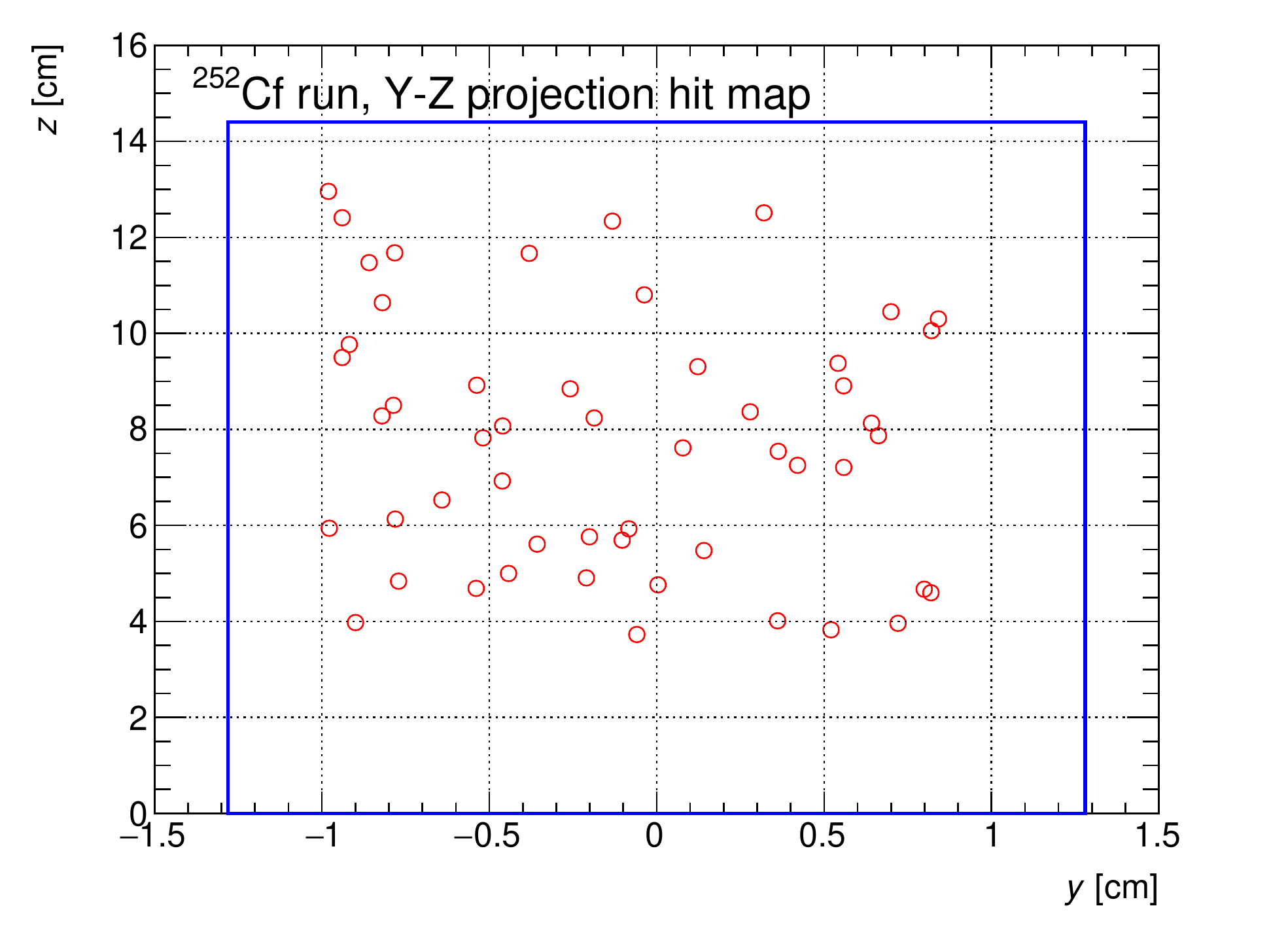}
					\hspace{1.6cm}
				\end{center}
			\end{minipage}
			\begin{minipage}{0.5 \hsize}
				\begin{center}
					\includegraphics[clip, width=7.4cm]{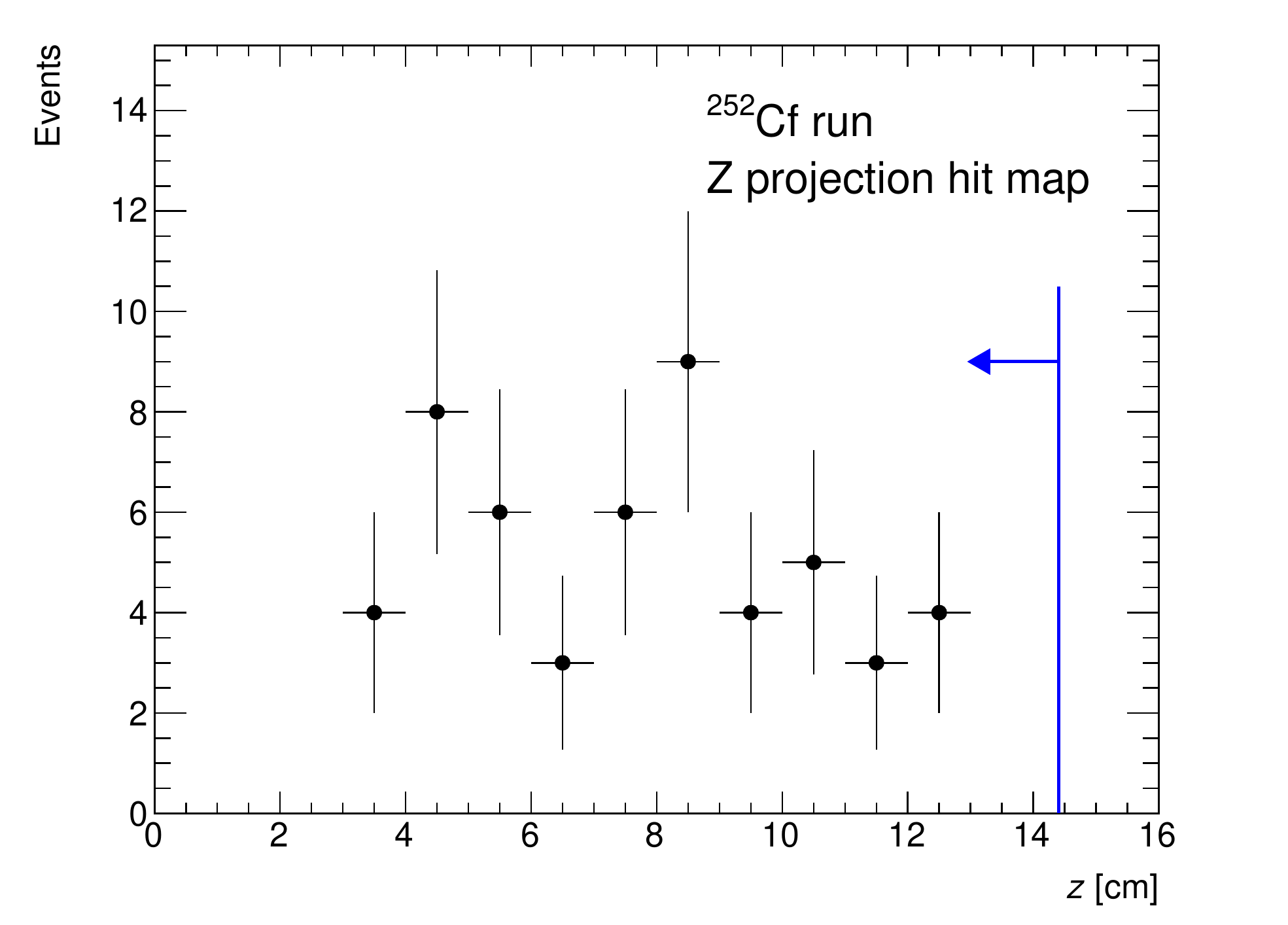}
					\hspace{1.6cm}
				\end{center}
			\end{minipage}
		\end{tabular}

		\caption{Distribution of absolute position of reconstructed tracks in the $^{252}$Cf run: $x-y$ (left top), $x-z$ (right top), $y-z$ (left bottom) and $z$ (right bottom) projections. Blue boxes and line represent the detection area. Events after all selections are plotted.}
		\label{fig:hitmap_projection}
	\end{center}
\end{figure}

\begin{figure}[h]
    \centering
    \includegraphics[width=0.85\textwidth]{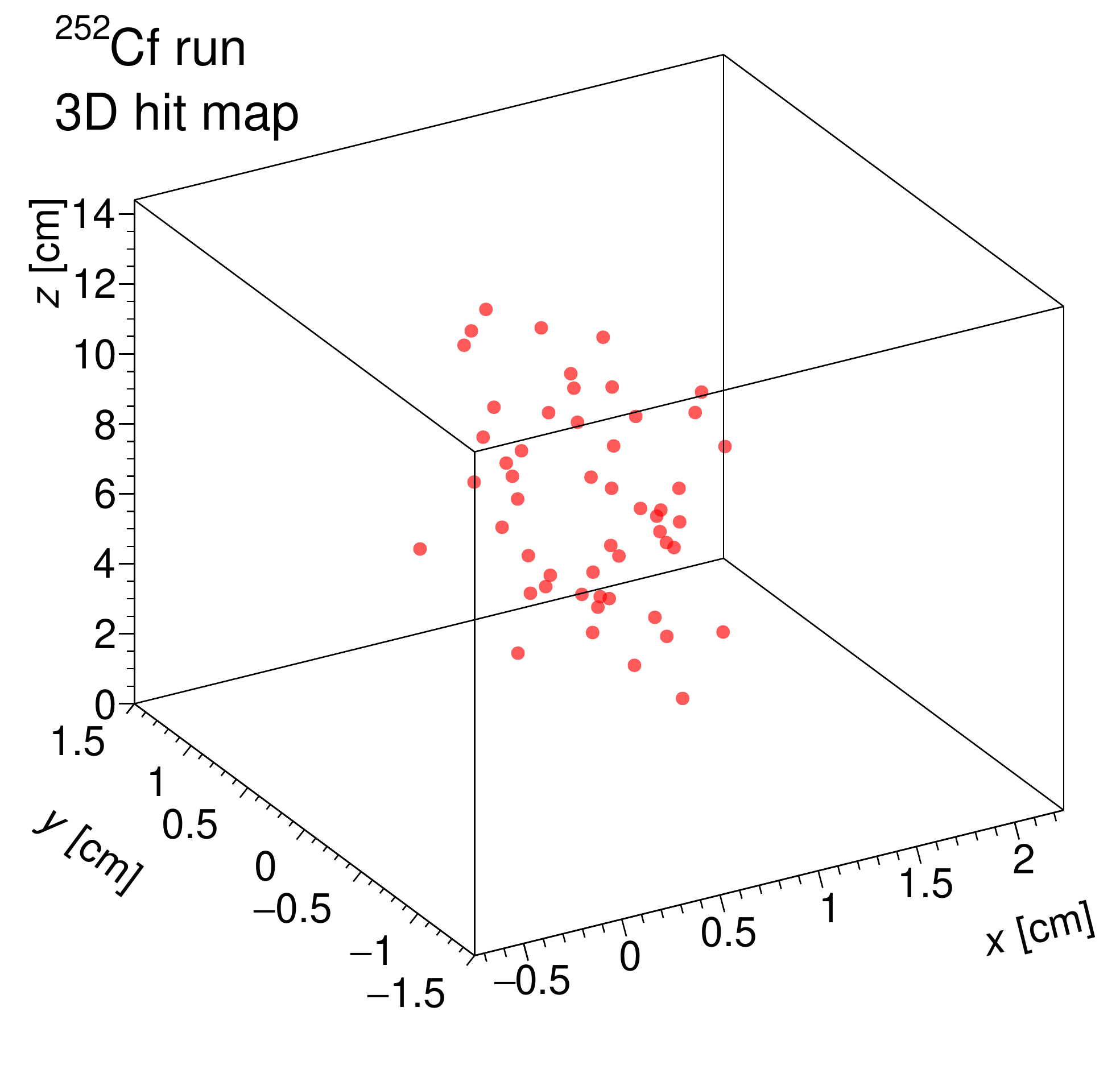}
    \caption{Distribution of three-dimensional absolute hit positions in $^{252}$Cf run after all selections.}
    \label{fig:3Dhitmap}
\end{figure}

The absolute position of the drift direction, defined as the z coordinate, was calculated by
\begin{equation}
    z = \frac{v_{{\rm SF}_{5}^{-}} \cdot v_{{\rm SF}_{6}^{-}}}{v_{{\rm SF}_{5}^{-}} - v_{{\rm SF}_{6}^{-}}} \Delta t,
\end{equation}
where $v_{{\rm SF}_{5}^{-}}$ and $v_{{\rm SF}_{6}^{-}}$ are the velocities of SF$_{5}^{-}$ and SF$_{6}^{-}$ carriers, respectively, and $\Delta t$ is the time difference between two peaks.
Figure~\ref{fig:hitmap_projection} shows the projected hit maps of absolute position of reconstructed tracks in the $^{252}$Cf run.
Events are distributed uniformly in the $x-y$ projection map.
Since the $^{252}$Cf source was placed at $z = 89$~mm, events reconstructed around $z = 89$~mm are slightly increased.
The event reconstruction is successfully working on any spacial absolute hit positions.
Although three events that are out of detection area along with $z$ axis were miss-reconstructed due to the position resolution, events were successfully reconstructed in the detection volume.
Figure~\ref{fig:3Dhitmap} demonstrates the three-dimensional absolute hit position reconstruction.
Here $x$ and $y$ axes are defined as the direction perpendicular to anode and cathode strips, respectively.
The three-dimensional hit position reconstruction was successfully working without biases in any hit position.

We hereby showed the first detection of three-dimensional absolute position reconstruction of $\mu$TPC using a negative ion gas.
In this study, the statistics was limited due to the small detection area.
In order to increase the size of the fiducial region, a mass-production of the electronic is required.
The extension of the fiducial region, equivalent to increase statistics, also enables to study the position and angular resolutions of tracks.
With these studies accomplished, a directional dark matter search with a NI$\mu$TPCs will be taken places.

\section{Conclusion}
\label{sec:conclusion}

We developed a prototype negative ion micro TPC using a $\mu$-PIC readout with dedicated electronics system filled with SF$_{6}$ gas at 20~torr.
We irradiated the detector with neutrons to emulate DM-induced nuclear recoil.
We succeeded in reconstructing three-dimensional absolute hit positions of nuclear recoil tracks with an efficiency of 70~$\pm$~5\%.
We also demonstrated to reconstruct tracks independent on its absolute position along with drift direction. These results mark an important step for the NI$\mu$TPCs towards the practical use for directional dark matter searches.

\section*{Acknowledgement}
\label{sec:ack}

This work was partially supported by KAKENHI Grant-in-Aids (16H02189, 19H05806, 21K13943 and 22H04574).

\vspace{0.2cm}
\noindent
\let\doi\relax
\bibliographystyle{unsrt}
\bibliography{ref}

\end{document}